\documentclass[aps,prd,onecolumn,groupedaddress,showpacs,nofootinbib,amssymb]{revtex4}
\usepackage{amsmath}
\usepackage{amssymb}
\usepackage{wrapfig}
\usepackage{cancel}

\newcommand{\e}{\mathrm{e}}
\usepackage{relsize,exscale}
\makeatletter

\usepackage{array, multirow,graphicx}
\usepackage{dcolumn}
\usepackage{newlfont}
\usepackage{bm}
\usepackage[colorlinks,citecolor=blue,urlcolor=blue,linkcolor=blue]{hyperref}
\usepackage[figtopcap]{subfigure}
\usepackage{color}

\allowdisplaybreaks[4]

\begin{document}
\tolerance=5000

\title{Hayward black hole in scalar-Einstein-Gauss-Bonnet gravity in four dimensions}

\author{Shin'ichi~Nojiri}
\email{nojiri@gravity.phys.nagoya-u.ac.jp}
\affiliation{Department of Physics, Nagoya University, Nagoya 464-8602, Japan \\
\& \\
Kobayashi-Maskawa Institute for the Origin of Particles and the Universe,
Nagoya University, Nagoya 464-8602, Japan }
\author{G.~G.~L.~Nashed}
\email{nashed@bue.edu.eg}
\affiliation {Centre for Theoretical Physics, The British University in Egypt, P.O. Box 43, El Sherouk City, Cairo 11837, Egypt}

\begin{abstract}

In the framework of scalar-Einstein-Gauss-Bonnet gravity, we construct the model which realizes the Hayward black hole and
discuss the absence of ghosts in this model. Because Hayward black hole has two horizons but no curvature singularity,
it may solve the problem of the information loss that might be generated by black holes.
The Gauss-Bonnet term appears as a stringy correction, and therefore, our results might indicate that the stringy correction would solve the information loss problem.
\end{abstract}


\maketitle

\section{Introduction}\label{S1}

Black holes (BHs) are extremely unusual regions of spacetime.
They are defined by the existence of the event horizon, which is a one-way causal spacetime boundary beyond which even light cannot escape.
BHs are understood to be the final state of the gravitational collapse of matter and are a generic prediction of
general relativity (GR) \cite{Einstein:1916vd,Schwarzschild:1916uq,Penrose:1964wq}.
Furthermore, BHs may hold the key to realizing the dream of unifying GR and quantum mechanics \cite{Hawking:1976ra, Giddings:2017jts} (see also \cite{Giddings:2019jwy}).
Undoubtedly, a deeper understanding of BHs will lead to a more complete understanding of gravity at energy scales that we cannot access from Earth.

 From an observational perspective, BHs can be seen in a wide range of astrophysical settings, and there is a plenitude of direct and indirect data suggesting
the existence of supermassive black holes (SMBHs) with masses as large as $10^{10}M_\odot$.
The SMBHs are thought to be located at the centers of the majority of galaxies with sufficiently large mass, including our galaxy~\cite{Lynden-Bell:1969gsv, Kormendy:1995er}.
They are thought to be a source of energy that often outstrips other galaxies, called the active galactic nucleus, where the center of the galaxy is extremely luminous.
For a comprehensive overview of astrophysical BHs, see \cite{Bambi:2019xzp}.

The gravitational collapse in GR would result in the appearance of singularities, which could be unavoidable and sometimes unwelcome~\cite{Penrose:1964wq, Hawking:1970zqf, Senovilla:1998oua}.
It would be desirable if we could find any solution that completely avoids the  singularities, even though the cosmic censorship conjecture predicts that all singularities
of gravitational collapse should be veiled beneath the event horizons of BHs and so should not be naked \cite{Penrose:1969pc, Wald:1997wa}.
Since Bardeen's early work \cite{bardeen1968non}, a lot of attention has been focused on finding regular BH solutions without singularity.
To do this, one can either alter the gravity sector or look for (usually exotic) matter content that can regularize the central singularity.
For an exhaustive list of works in this area, see~\cite{Borde:1996df, Ayon-Beato:1998hmi, Ayon-Beato:1999kuh, Bronnikov:2000yz, Burgess:2005sb, Hayward:2005gi,
Bronnikov:2005gm,Iso:2006ut,Berej:2006cc,Li:2008fa,DeFelice:2009rw,Bronnikov:2012ch,Rinaldi:2012vy,Bambi:2013ufa,Sebastiani:2013fsa,Toshmatov:2014nya,
Johannsen:2013szh,Myrzakulov:2015sea,Myrzakulov:2015qaa,Myrzakulov:2015kda,Abdujabbarov:2016hnw,Fan:2016hvf,Sebastiani:2016ras,Toshmatov:2017zpr,
Chinaglia:2017uqd, Chinaglia:2017wih, Colleaux:2017ibe, Jusufi:2018jof, Chinaglia:2018gvf, Ovgun:2019wej,Han:2019lfs,Rodrigues:2019xrc,Panotopoulos:2019qjk,Jusufi:2019caq,Gorji:2020ten}.
For significant investigations on the observational signatures of such BHs, see \cite{Schee:2015nua, Stuchlik:2014qja, Schee:2016mjd, Stuchlik:2019uvf, Schee:2019gki}.

One of the most important problems in quantum gravity could be the so-called ``information loss problem'' by BH evaporation.
Hawking radiation~\cite{Hawking:1975vcx} seems to tell us that the initial quantum pure state, which describes gravitationally collapsing matter to form a BH,
evolves into the quantum mixed state, and therefore the unitarity, which is one of the basic principles in quantum theory, seems to be broken.
There have been several proposed scenarios to solve this problem of the loss of quantum information.
Hawking himself has proposed that the black hole completely evaporates and the quantum coherence could be lost and the unitarity is broken.
On the other hand, 't Hooft has considered that Hawking radiation may carry information about the quantum states of the black holes~\cite{tHooft:1990fkf}.
There is also a proposal that after Hawking radiation, there remains a stable remnant that carries the information of the initial state~\cite{Aharonov:1987tp}.
A candidate of the remnant is the extremal limit of the Reissner-Nordstr\"om BHs~\cite{Holzhey:1991bx}.
More recently, the importance of infrared physics has been pointed out.
Soft hairs associated with the asymptotic symmetries in the infrared region may store the information before the collapse to the BH~\cite{Hawking:2016msc}

Another scenario could be given by the regular BH where any curvature singularity does not appear.
In a semi-classical description of Hawking radiation, there occurs the pair creation of two particles near the horizon.
One of the particles escapes to infinity and the particle is recognized as radiation.
Another particle falls into the singularity of the BH, and the information is lost.
If there is no singularity, the fallen particle goes out through the horizons, again, and arrives at another universe
and the information is carried to the universe.
Therefore, if we include another universe, information loss does not occur.
Such a regular BH has been constructed by using non-linear electrodynamics~\cite{Bardeen1968} (see also \cite{Nojiri:2017kex}).
Hayward also proposed the metric of the regular BH~\cite{Hayward:2005gi} but the gravity theory which realizes Hayward BH was not given.
In this paper, in the framework of the scalar-Einstein-Gauss-Bonnet gravity,
we propose a model which realizes Hayward BH although Hayward BH is also constructed by using the non-linear electrodynamics \cite{Fan:2016hvf}.

The reason why we consider the Einstein-Gauss-Bonnet gravity is that the gravity theory appears as a correction coming
from the string theory~\cite{Boulware:1985wk}.
String theory is the strongest candidate for quantum gravity and it could be natural to expect that quantum gravity could solve
the problem of information loss.
Because the Gauss-Bonnet invariant is a topological quantity, that is, total derivative in four dimensions, the BH solutions have been well
studied in dimensions equal to or higher than five \cite{Wiltshire:1985us} (see also \cite{Cvetic:2001bk}).
If we consider the dimensional reduction from the higher dimensions to four dimensions, there could appear the coupling between the
Gauss-Bonnet term and other fields.
The simplest model of this type is the scalar-Einstein-Gauss-Bonnet gravity, which we consider in this paper.
In cosmology, the scalar-Einstein-Gauss-Bonnet gravity has been also used to explain dark energy, and it has been shown
that the phantom universe is realized without ghosts~\cite{Nojiri:2005vv}.
Furthermore, it has been found that arbitrary development of the expansion of the universe can be realized
in the scalar-Einstein-Gauss-Bonnet gravity~\cite{Nojiri:2006je}.
Even for BH-like objects, the authors have found that arbitrary spherically symmetric and static spacetime can be constructed
in the framework of scalar Einstein-Gauss-Bonnet gravity, although there might appear ghosts~\cite{Nashed:2021cfs}.
In this paper, by using the formulation in \cite{Nashed:2021cfs}, we construct the model which realizes Hayward BH and discusses
the possibility of the absence of ghosts.

In the next section, we give the equations in the scalar-Einstein-Gauss-Bonnet gravity model, and we show how to construct a model
which realizes general spherically symmetric and static solutions.
In Section~\ref{S4}, after briefly reviewing the properties of Hayward BH, we construct a model that realizes Hayward BH.
We especially discuss how we could be able to avoid ghosts.
We also investigate the solution's small- and large-$r$ behaviors for general spherically symmetric and static solutions and check the obtained results.
In Section~\ref{S66}, by using the formulation of the geodesic deviation, the stability condition of a particle motion in the background of Hayward BH is studied.
The last section is devoted to the summary and discussions.

\section{General spherically symmetric and static solution of scalar-Einstein-Gauss-Bonnet gravity in four dimensions}\label{S3}

The action of the scalar-Einstein-Gauss-Bonnet gravity in $N$ spacetime dimensions is given by,
\begin{align}
\label{g2}
\mathcal{S}=\int d^N x \sqrt{-g}\left\{ \frac{1}{2\kappa^2}R
 - \frac{1}{2} \partial_\mu \xi \partial^\mu \xi+V(\xi)+ f(\xi) \mathcal{G} \right\}\, .
\end{align}
Here $\xi$ is the scalar field and $V(\xi)$ is the potential for $\xi$ and $f(\xi)$ is also a function of $\xi$.
Furthermore, $\mathcal{G}$ is the Gauss-Bonnet invariant defined by
\begin{align}
\label{eq:GB}
\mathcal{G} = R^2-4R_{\alpha \beta}R^{\alpha \beta}+R_{\alpha \beta \rho \sigma}R^{\alpha \beta \rho \sigma}\, ,
\end{align}
Although the Gauss-Bonnet invariant is total-derivative in four dimensions ($N=4$), due to the coupling $f(\xi)$, the term including
the Gauss-Bonnet invariant gives non-trivial contributions to the field equations of the system.
{ We also assume that the matters do not couple with the scalar field $\xi$,
which may avoid the appearance of the fifth force.}

The variation of the action (\ref{g2}) with respect to the scalar field $\xi$ yields the following equation,
\begin{align}
\label{g3}
\nabla^2 \xi-V'(\xi)+ f'(\xi)\mathcal{G}=0\, .
\end{align}
On the other hand, by the variation of the action (\ref{g2}) with respect to the metric $g_{\mu\nu}$, we obtain the following equations,
\begin{align}
\label{GBeq}
0=&\, \frac{1}{2\kappa^2}\left(- R^{\mu\nu} + \frac{1}{2} g^{\mu\nu} R\right)+\frac{1}{2} \partial^\mu \xi \partial^\nu \xi
 - \frac{1}{4}g^{\mu\nu} \partial_\rho \xi \partial^\rho \xi+ \frac{1}{2} g^{\mu\nu}[f(\xi)G -V(\xi)]
+2 f(\xi) R R^{\mu\nu} \nonumber \\
&\, + 2 \nabla^\mu \nabla^\nu \left(f(\xi)R\right)- 2 g^{\mu\nu}\nabla^2\left(f(\xi)R\right)
+ 8f(\xi)R^\mu_{\ \rho} R^{\nu\rho}- 4 \nabla_\rho \nabla^\mu \left(f(\xi)R^{\nu\rho}\right)
 - 4 \nabla_\rho \nabla^\nu \left(f(\xi)R^{\mu\rho}\right) \nonumber \\
&\, + 4 \nabla^2 \left( f(\xi) R^{\mu\nu} \right)+ 4g^{\mu\nu} \nabla_{\rho} \nabla_\sigma \left(f(\xi) R^{\rho\sigma} \right)
 - 2 f(\xi) R^{\mu\rho\sigma\tau}R^\nu_{\ \rho\sigma\tau}+ 4 \nabla_\rho \nabla_\sigma \left(f(\xi) R^{\mu\rho\sigma\nu}\right)\, .
\end{align}
Using the Bianchi identities,
\begin{align}
\label{Bianchi}
\nabla^\rho R_{\rho\tau\mu\nu}=&\, \nabla_\mu R_{\nu\tau} - \nabla_\nu R_{\mu\tau} \, , \nonumber \\
\nabla^\rho R_{\rho\mu} =&\, \frac{1}{2} \nabla_\mu R\, , \nonumber \\
\nabla_\rho \nabla_\sigma R^{\mu\rho\nu\sigma} =&\, \nabla^2 R^{\mu\nu} - \frac{1}{2}\nabla^\mu \nabla^\nu R
+ R^{\mu\rho\nu\sigma} R_{\rho\sigma}- R^\mu_{\ \rho} R^{\nu\rho}\, , \nonumber \\
\nabla_\rho \nabla^\mu R^{\rho\nu} + \nabla_\rho \nabla^\nu R^{\rho\mu}
=&\, \frac{1}{2} \left(\nabla^\mu \nabla^\nu R + \nabla^\nu \nabla^\mu R\right)
 - 2 R^{\mu\rho\nu\sigma} R_{\rho\sigma} + 2 R^\mu_{\ \rho} R^{\nu\rho}\, , \nonumber \\
\nabla_\rho \nabla_\sigma R^{\rho\sigma} =& \frac{1}{2} \Box R \, ,
\end{align}
in Eq.~(\ref{GBeq}), we obtain
\begin{align}
\label{gb4b}
0=&\, \frac{1}{2\kappa^2}\left(- R^{\mu\nu} + \frac{1}{2} g^{\mu\nu} R\right)
+ \left(\frac{1}{2} \partial^\mu \xi \partial^\nu \xi
 - \frac{1}{4}g^{\mu\nu} \partial_\rho \xi \partial^\rho \xi \right)
+ \frac{1}{2} g^{\mu\nu} \left[ f(\xi) G-V(\xi) \right] \nonumber \\
&\, -2 f(\xi) R R^{\mu\nu} + 4f(\xi)R^\mu_{\ \rho} R^{\nu\rho}
 -2 f(\xi) R^{\mu\rho\sigma\tau}R^\nu_{\ \rho\sigma\tau}
 -4 f(\xi) R^{\mu\rho\sigma\nu}R_{\rho\sigma} \nonumber \\
&\, + 2 \left( \nabla^\mu \nabla^\nu f(\xi)\right)R
 - 2 g^{\mu\nu} \left( \nabla^2f(\xi)\right)R
 - 4 \left( \nabla_\rho \nabla^\mu f(\xi)\right)R^{\nu\rho}
 - 4 \left( \nabla_\rho \nabla^\nu f(\xi)\right)R^{\mu\rho} \nonumber \\
&\, + 4 \left( \nabla^2 f(\xi) \right)R^{\mu\nu}
+ 4g^{\mu\nu} \left( \nabla_{\rho} \nabla_\sigma f(\xi) \right) R^{\rho\sigma}
 - 4 \left(\nabla_\rho \nabla_\sigma f(\xi) \right) R^{\mu\rho\nu\sigma}.
\end{align}
In the 4-dimension case i.e., $N=4$, Eq.~(\ref{gb4b}) is reduced to
\begin{align}
\label{gb4bD4}
0= & \frac{1}{2\kappa^2}\left(- R^{\mu\nu} + \frac{1}{2} g^{\mu\nu} R\right)
+ \left(\frac{1}{2} \partial^\mu \xi \partial^\nu \xi
 - \frac{1}{4}g^{\mu\nu} \partial_\rho \xi \partial^\rho \xi \right)
 - \frac{1}{2} g^{\mu\nu}V(\xi) \nonumber \\
& + 2 \left( \nabla^\mu \nabla^\nu f(\xi)\right)R
 - 2 g^{\mu\nu} \left( \nabla^2f(\xi)\right)R
 - 4 \left( \nabla_\rho \nabla^\mu f(\xi)\right)R^{\nu\rho}
 - 4 \left( \nabla_\rho \nabla^\nu f(\xi)\right)R^{\mu\rho} \nonumber \\
& + 4 \left( \nabla^2 f(\xi) \right)R^{\mu\nu}
+ 4g^{\mu\nu} \left( \nabla_{\rho} \nabla_\sigma f(\xi) \right) R^{\rho\sigma}
- 4 \left(\nabla_\rho \nabla_\sigma f(\xi) \right) R^{\mu\rho\nu\sigma}.
\end{align}
This is because the Gauss-Bonnet invariant is a total derivative in four dimensions and therefore if $f$ is a constant,
the term including the Gauss-Bonnet invariant does not give any contribution to the equations in (\ref{gb4b}) or (\ref{gb4bD4}),
which tells that any term including $f(\xi)$ without derivative of $f(\xi)$ should vanish in four-dimensions.

In the present study, we consider the following spherically symmetric and static spacetime,
\begin{align}
\label{met1}
ds^2 = -a(r)dt^2 +\frac{dr^2}{a(r)}+r^2 \left( d\theta^2 + \sin^2 \left(\theta\right)\right) d\phi^2\,.
\end{align}
Therefore we may assume that the scalar field $\xi$ is a function of only the radial coordinate $r$ and therefore the function $f(\xi)$ is also a function of $r$, i.e.,
$f(r)\equiv f\left(\xi\left(r\right) \right)$.

Under the metric given by Eq. (\ref{met1}), the field equations (\ref{gb4bD4}) take the following forms,
\begin{align}
\label{Eq2tt}
0=&\, \frac{16a\left(1 -a\right) f'' + \left\{ 8\left( 1-3a \right)f' +2r \right\} a'_1
+2a +r^2 \xi'^2 a -2+2r^2V}{4r^2} \,, \\
\label{Eq2rr}
0=&\, \frac{2\left(4 \left(1 -3a \right) f' +r \right)a'+2a -r^2 a\xi'^2 -2 +2V r^2 }{4 r^2 }\,, \\
\label{Eq2pp}
0=&\, \frac{\left(r - 8 f' a \right)a'' -8 f''a a' - 8 f' {a'}^2 + 2 a' +r\left(2V+ \xi'^2a \right)}{4r} \,.
\end{align}
Here the prime $'$ expresses the derivative with respect to $r$, for example, $a'=\frac{da}{dr}$.
Eqs.~(\ref{Eq2tt}) and (\ref{Eq2rr}) are the $(t,t)$-component and $(r,r)$ -componet of Eq.~(\ref{gb4bD4}), respectively.
On the other hand the $(\theta,\theta)$ and $(\phi,\phi)$-components give the identical equation~(\ref{Eq2pp}).

The equation of the scalar field $\xi$ in (\ref{g3}) has the following form,
\begin{align}
\label{Eqphi2}
0=\frac{4 f' \left(a-1 \right) a'' + \xi'' a \xi'r^2 +4 a'^2f' +r \left[ a' r+2a \right]\xi'^2-r^2V' }{ r^2 \xi'} \,.
\end{align}
By combining Eq.~(\ref{Eq2tt}) and Eq.~(\ref{Eq2rr}), we obtain
\begin{align}
\label{V2}
V = \frac{1-4 a \left( 1-a \right) f'' - \left[ 4a' \left( 1-3a \right) f'+r \right] a'-a }{r^2} \, .
\end{align}
Furthermore, Eq.~(\ref{Eq2tt}) and Eq.~(\ref{Eq2rr}) yields
\begin{align}
\label{xi3}
\xi'=\frac{2}{r}\sqrt{2 \left( a-1 \right)f''}\,.
\end{align}
Finally, by using Eq.~(\ref{Eq2tt}) and Eq.~(\ref{Eq2pp}), we find
\begin{align}
\label{f}
0= 8a \left( 2a-a'r-2 \right)f''+a''r \left( r-8f'a \right)-8rf'a'^2+8 \left( 3a-1 \right)f'a'+2 \left( 1-a \right) \,.
\end{align}
Eq.~(\ref{xi3}) tells if $\left( a-1 \right)f''<0$, the scalar field $\xi$ becomes imaginary number.
In this case, by redefining the scalar field $\xi=i\zeta$, we obtain a real scalar field $\zeta$ but for the redefined
scalar field $\zeta$, the signature of the kinetic term in (\ref{g2}) is changed
$ - \frac{1}{2} \partial_\mu \xi \partial^\mu \xi = + \frac{1}{2} \partial_\mu \zeta \partial^\mu \zeta$.
This tells that the scalar field $\zeta$ is a ghost and therefore the system is physically inconsistent.
Classically the ghost generates the unbounded negative kinetic energy and in quantum theory, there appear negative norm states.
The negative norm states which generate the negative probability conflict with the Copenhagen ansatz, which is one of the basic assumptions in
the quantum theory.

We should note that Eq.~(\ref{f}) can be trivially integrated and we obtain,
\begin{align}
\label{f1}
f(r)=-\frac{1}{8}\int \left( \int \frac{\e^{\int \frac{a' -3 a a' +r{a'}^2+ra a'' }{a \left( 2-2\,a + a' r \right)}dr}
\left(2 a-2 - a'' r^2 \right) }{ a\left( 2-2a + a' r \right)}dr -8 c_0 \right) { \e^{-\int \frac{a' -3 a a' +r {a'}^2+ra a'' }{a \left( 2-2 a + a' r \right) }dr}} dr+c_1\,,
\end{align}
where $c_0$ and $c_1$ are constants of integration.

When $a=a(r)$ in (\ref{met1}) is given, we find the $r$-dependence of $f$, i.e., $f=f(r)$ by using (\ref{f1}).
Then by using (\ref{V2}), we find the $r$-dependence of the potential $V=V(r)$ and furthermore by using (\ref{xi3}), we also obtain
the $r$-dependence of the scalar field $\xi=\xi(r)$, which could be solved with respect to $\xi$, $r=r(\xi)$.
Because we find the $r$-dependence of $f=f(r)$ and $V=V(r)$, if we delete $r$ by using $r=r(\xi)$,
we obtain $f$ and $V$ as functions of $\xi$, $V(\xi)=V\left(r\left(\xi\right)\right)$ and $f(\xi)=f\left(r\left(\xi\right)\right)$,
which gives the model to realize the metric in (\ref{met1}).

\section{Hayward BH}\label{S4}

{
\subsection{Properties}
}

In this paper, we construct Hayward BH~\cite{Hayward:2005gi} as a solution of the scalar-Einstein-Gauss-Bonnet gravity.
For this purpose, in this section, we start to describe the properties of Hayward BH.

The metric of the spacetime describing Hayward BH~\cite{Hayward:2005gi} is given by
\begin{align}
\label{Hayward1}
ds^2 = -a(r)dt^2 +\frac{dr^2}{a(r)}+r^2 \left( d\theta^2 + \sin^2 \left(\theta\right)\right) d\phi^2\,, \qquad
a(r)= 1 - \frac{2Mr^2}{r^3 + 2M\lambda^2} \, .
\end{align}
We should note that $a'(r)$ vanishes at the center $r=0$, $a'(r=0)=0$, and therefore, there is no conical singularity.
In fact, when $r$ is small, we find $a(r) \sim 1 - \frac{r^2}{\lambda^2}$, which gives a metric of the de Sitter spacetime in a static patch.

We rewrite $a(r)$ in (\ref{Hayward1}) as follows,
\begin{align}
\label{Hayward2}
a(r) = \frac{b(r)}{r^3 + 2M\lambda^2} \, , \qquad \qquad b(r)\equiv r^3 - 2Mr^2 + 2M\lambda^2 \, .
\end{align}
Then the solutions of $b'(r) = 3r^2 - 4Mr=0$ are given by $r=0$ and $r=\frac{4}{3}M$.
Because
\begin{align}
\label{Hayward3}
b\left( r =\frac{4}{3}M \right) = - \frac{2^5}{3^3} M^3 + 2M\lambda^2 \, ,
\end{align}
we find
\begin{enumerate}
\item In the case that the following inequality is satisfied,
\begin{align}
\label{condition1}
\frac{2^\frac{5}{3}M}{3\left( 2M\lambda^2 \right)^\frac{1}{3}}<1\, ,
\end{align}
$a$ is always positive and therefore the metric (\ref{Hayward1}) describes a kind of the gravasar.
Gravastar was proposed in \cite{Mazur:2001fv} as an alternative to the BH.
In the original proposal, the gravastar has the usual BH metric outside of the horizon but the de Sitter metric inside the horizon and
on the horizon, there is a thin shell of matter.
In the metric of (\ref{Hayward1}), when $r$ is large, $a(r)$ behaves as a BH $a(r)\sim 1 - \frac{2M} r$ but there is no singularity at the center
and behaves as the de Sitter spacetime.
Therefore the metric (\ref{Hayward1}) surely describes the gravastar without a shell when the condition (\ref{condition1}) is satisfied.
\item On the other hand, when
\begin{align}
\label{condition2}
\frac{2^\frac{5}{3}M}{3\left( 2M\lambda^2 \right)^\frac{1}{3}}>1\, ,
\end{align}
$a(r)$ has two zeros, which correspond to the outer and inner horizons, and therefore the metric corresponds to the regular BH in \cite{Hayward:2005gi} which is shown
in Fig.~\ref{Fig:1}~\subref{fig:1a}.
\begin{figure}
\subfigure[~The plot of the function $a(r)$, given by Eq.~\eqref{Hayward2}, vs. the radial coordinate $r$ which creates the inner and outer horizons ]{\label{fig:1a}\includegraphics[scale=0.25]{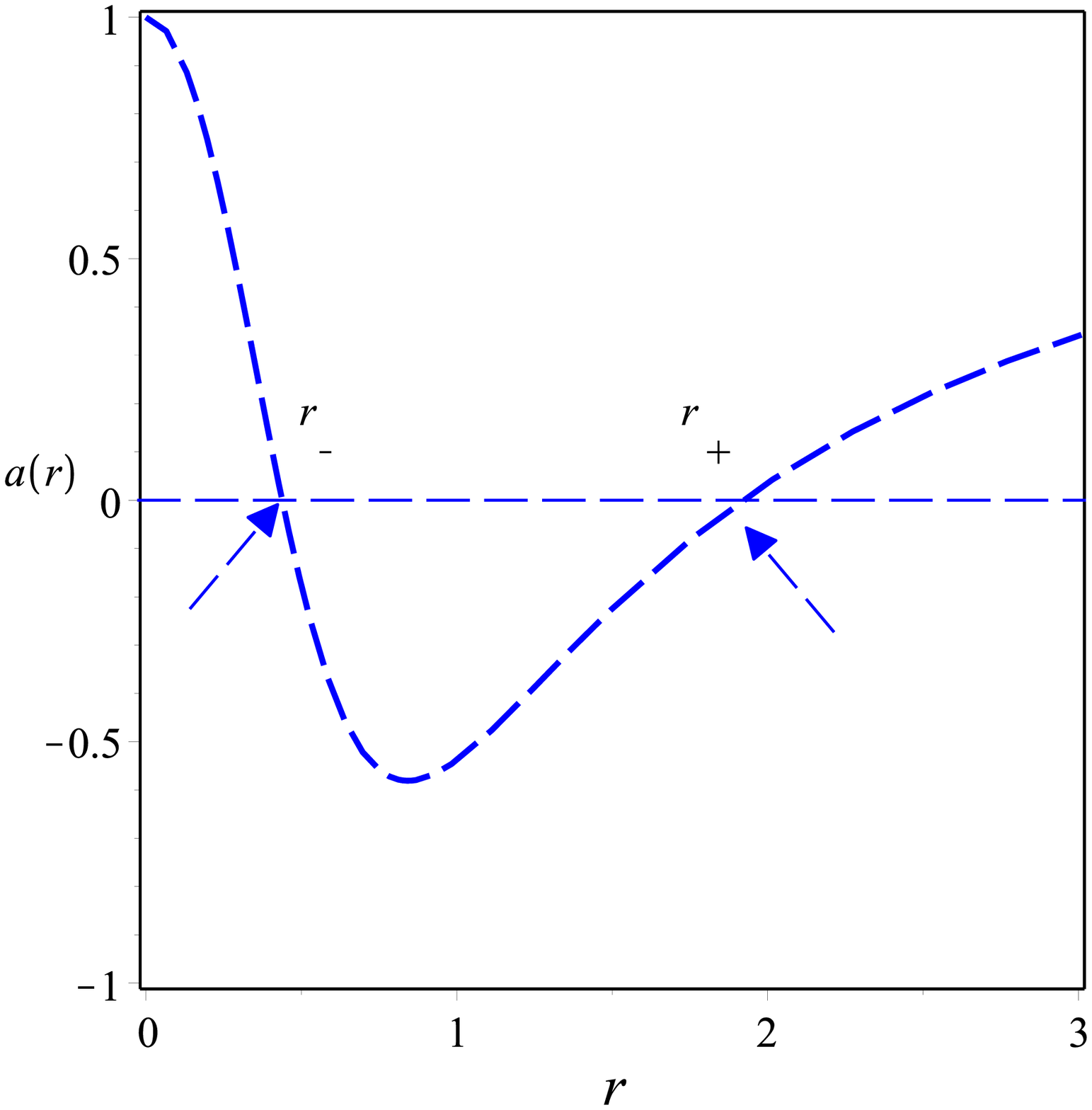}}\hspace{0.2cm}
\subfigure[~The plot of the function $a(r)$, given by Eq.~\eqref{Hayward2}, vs. the radial coordinate $r$ which creates the degenerate horizon ]{\label{fig:1b}\includegraphics[scale=0.25]{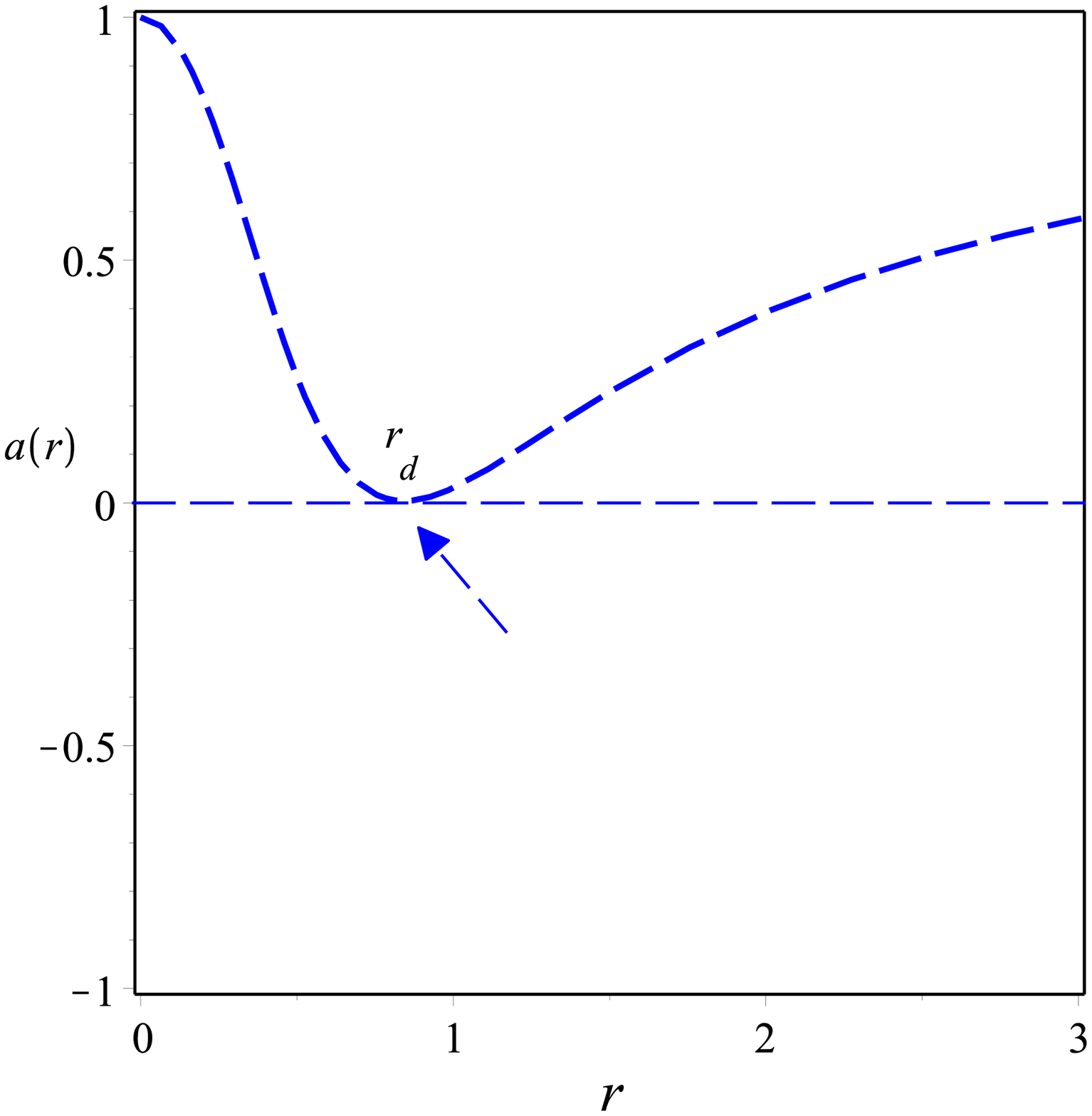}}\hspace{0.2cm}
\subfigure[~The plot of the temperature, given by Eq.~\eqref{Hayward6}, vs. the inner horizon ]{\label{fig:1c}\includegraphics[scale=0.25]{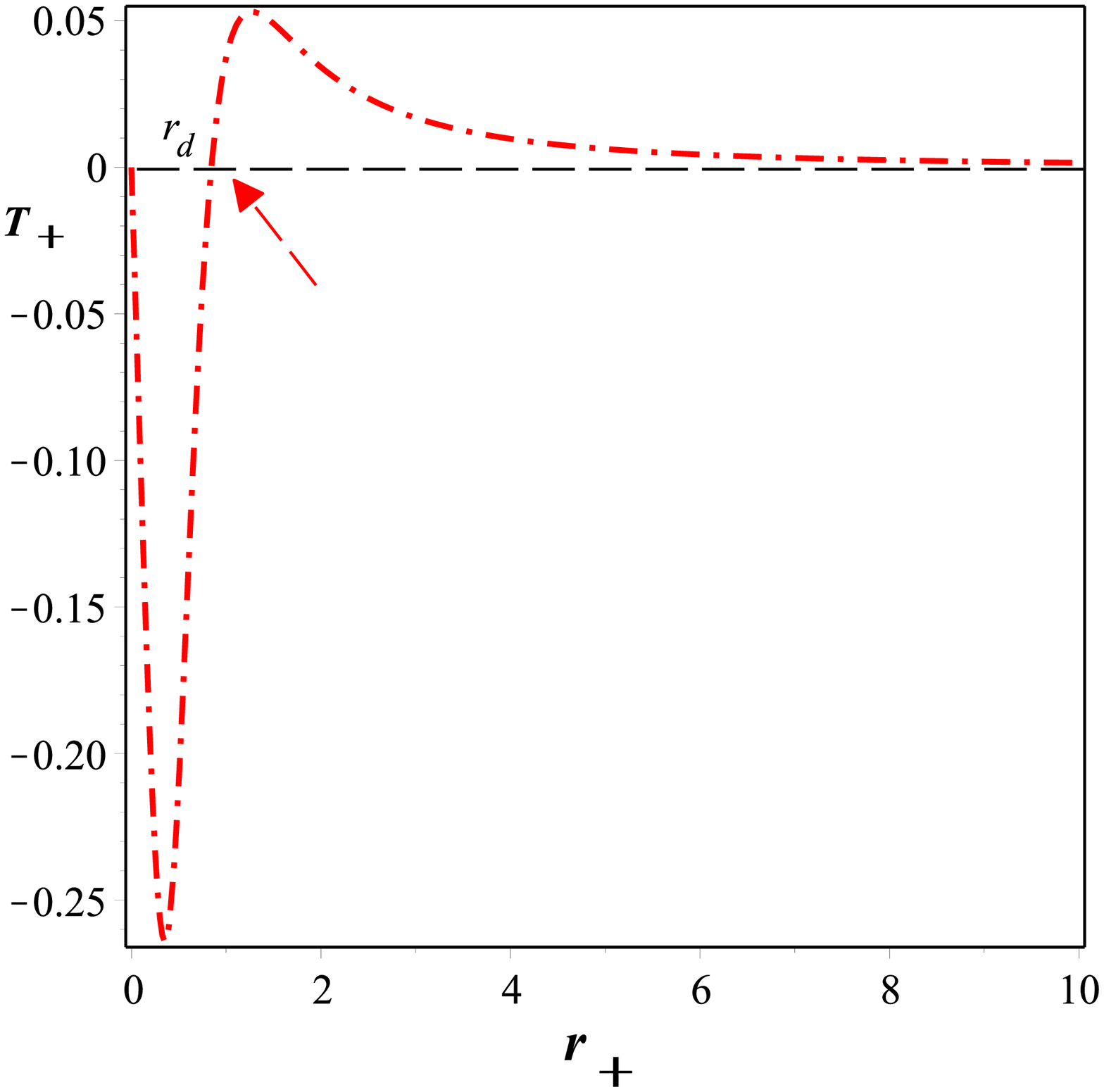}}\hspace{0.2cm}
\caption{
Schematic plot of the radial coordinate $r$ vs. the function $a$ \subref{fig:1a} gives the two horizons where we have put $M=1$ and $\lambda=0.3$;
\subref{fig:1b} the degenerate horizon where we have put $M={ 0.8}$ and $\lambda=0.3$;
\subref{fig:1c} the Hawking temperature where we have put $M=1$ and $\lambda=0.3$.}
\label{Fig:1}
\end{figure}

Because there is no singularity at the center, the object fallen into the BH beyond the outer and inner horizon can escape the BH by going through
the inner and outer horizons, again.
The Penrose diagram of this spacetime is given in the original paper by Hayward \cite{Hayward:2005gi}.
The Penrose diagram tells that the universe before the object falls into BH is different from the universe in which the object escapes.
In this sense, the regular BH plays the role of the wormhole.
\item When
\begin{align}
\label{condition3}
\frac{2^\frac{5}{3}M}{3\left( 2M\lambda^2 \right)^\frac{1}{3}}=1\, ,
\end{align}
the radii of the two horizons become identical with each other and constitute the degenerate horizon and therefore it corresponds to the extremal limit of the BH which is shown in Fig.~\ref{Fig:1}~\subref{fig:1b}.
\end{enumerate}

If the condition~(\ref{condition2}) is satisfied, $a(r)$ can be rewritten as follows:
\begin{align}
\label{Hayward4}
a(r) =&\, \frac{\left( r - r_+\right) \left( r - r_-\right) \left( r + \frac{r_+ r_-}{r_+ + r_-}\right)}{r^3 + 2M\lambda^2} \, ,\qquad \qquad
2M = \frac{{r_+}^2 + {r_-}^2 + r_+ r_-}{r_+ + r_-} \, , \qquad \qquad
2M\lambda^2 = \frac{{r_+}^2 {r_-}^2}{r_+ + r_-} \, .
\end{align}
or
\begin{align}
\label{rpm}
r_\pm =\frac{M^\frac{2}{3}\left[ \left\{8M^2-27\lambda^2 \pm 3\lambda\sqrt{81\lambda^2-48M^2} \right\}^\frac{1}{3}
+2M^\frac{2}{3} \right] \left\{8M^2-27\lambda^2 \pm 3\lambda\sqrt{81\lambda^2-48M^2} \right\}^\frac{1}{3} +4M^2}
{3M^\frac{1}{3} \left\{ 8M^2-27\lambda^2 \pm 3\lambda\sqrt{81\lambda^2-48M^2} \right\}^\frac{1}{3} }\,.
\end{align}
We should note that $r_+>r_->0$.
The surface gravities $\kappa_\pm$ on the horizons are defined by
\begin{align}
\label{Hayward5}
\kappa_\pm \equiv \left| \frac{1}{2}\left. a'(r) \right|_{r=r_\pm} \right|
= \frac{\left( r_+ - r_-\right) \left( r_\pm + \frac{r_+ r_-}{r_+ + r_-}\right)}{2\left({r_\pm}^3 + 2M\lambda^2\right)}
= \frac{\left( r_+ - r_-\right) \left( {r_\pm}^2 + 2 r_+ r_- \right)}{2\left({r_\pm}^3 + 2M\lambda^2\right)\left( r_+ + r_- \right) } \, ,
\end{align}
and the Hawking temperatures $T_\pm$ are given by
\begin{align}
\label{Hayward6}
T_\pm = \frac{\kappa_\pm}{2\pi} \equiv \left| \frac{1}{2}\left. a'(r) \right|_{r=r_\pm} \right|
= \frac{\left( r_+ - r_-\right) \left( {r_\pm}^2 + 2 r_+ r_- \right)}{4\pi \left({r_\pm}^3 + 2M\lambda^2\right)\left( r_+ + r_- \right) } \, ,
\end{align}
We should note that in the extremal limit where the radii of the horizons coincide with each other, $r_+=r_-=r_d$,
the Hawking temperature vanishes as in the Reissner-Nordstr\"om BH.
In Fig.~\ref{Fig:1}~\subref{fig:1c}, we show the behavior of the Hawking temperature which indicates where the two horizons coincide with each other.

{

\subsection{A solution of information loss problem}

The standard arguments for the information loss by the black hole are the following:
Let us suppose the standard black hole is formed by the collapse of matters.
After the formation of the black hole, the Hawking radiation could occur.
Because the black hole is characterized only by three hairs, that are, the mass,
the angular momentum, and the electric charge, the thermal radiation reflect only the three pieces of information,
and therefore most of the information in the collapsing matters is lost if the black hole evaporates by the Hawking radiation.

In the case of a regular black hole like the Hayward one, after the collapsing matters go through the inner horizon,
because there is no singularity, the matters may go through the horizons, again, and go out to the (possibly another) space-time.
The Hawking radiation could be the pair creations of particles near the horizon and one particle goes out to infinity and is observed as radiation.
Another particle falls into the black hole and in the standard black hole with singularity, the particle arrives at the singularity
and the information carried by the falling particle is lost.
In the case of the regular black hole, the falling particle may appear in (another) space-time by going through the horizons.
The falling particle carries also the information of the outgoing particle which may be observed as the Hawking radiation.
Therefore if we include the (another) space-time, the total information is not lost.

\subsection{Realization of Hayward BH}\label{S5}
}

Now, we are going to construct a model of the scalar-Einstein-Gauss-Bonnet gravity, which realizes Hayward BH in (\ref{Hayward1}).
Because
\begin{align}
\label{Hayward15}
a' = - \frac{- 2 Mr^4 + 8M^2 \lambda^2 r}{ \left( r^3 + 2M\lambda^2 \right)^2} \, , \qquad \qquad
a'' = - \frac{4Mr^6 - 56 M^2 \lambda^2 r^3 + 16 M^3 \lambda^4}{ \left( r^3 + 2M\lambda^2 \right)^3 } \, .
\end{align}
Eq.~(\ref{f1}) has the following form,
\begin{align}
\label{f3}
f(r)=&\, \frac{1}{2}\int \left( 3M{\lambda}^2\int \frac{\e^{4\int \Upsilon(r)dr}}{ \left( r^3+2 M{\lambda}^2-2M r^2 \right) }{dr}
+2c_0 \right) { \e^{-4\int\Upsilon(r){dr} }}{dr}+c_1\,, \nonumber \\
\Upsilon(r)=&\, \frac{6 r^6M{\lambda}^2- r^9+18 r^3 M^2{\lambda}^{4}+4 M^3{\lambda}^6+3M r^8-21 M^2 r^5{\lambda}^2}{3r^4
\left( r^3+2M{\lambda}^2-2M r^2 \right) \left( r^3+2M{\lambda}^2 \right) } \,.
\end{align}
Then, by using Eqs.~(\ref{V2}) and (\ref{xi3}), we find the $r$-dependence of the potential $V=V(r)$ and the scalar field $\xi=\xi(r)$
We may solve the equation $\xi=\xi(r)$ with respect to $\xi$, $r=r(\xi)$.
By deleting $r$ in $f=f(r)$ and $V=V(r)$ by using $r=r(\xi)$,
we obtain $f$ and $V$ as functions of $\xi$, $V(\xi)=V\left(r\left(\xi\right)\right)$ and $f(\xi)=f\left(r\left(\xi\right)\right)$,
which gives the model to realize Hayward BH (\ref{Hayward1}).

{
\subsection{Avoiding ghosts}

We now check if we can avoid ghosts.
}

When $r$ is small, we find
\begin{align}
\label{Hywrd1}
\Upsilon(r) \sim \frac{M\lambda^2}{3r^4} \, , \qquad
f(r) \sim \int \left( \frac{3}{4} \int \e^{- \frac{4M\lambda^2}{9r^3}} dr+ c_0 \right) \e^{ \frac{4M\lambda^2}{9r^3}} dr + c_1 \, ,
\end{align}
Because
\begin{align}
\label{Hywrd2}
\int \e^{- \frac{4M\lambda^2}{9r^3}} dr
=&\, \int \frac{3r^4}{4M\lambda^2}\left( \frac{4M\lambda^2}{3r^4} \right) \e^{- \frac{4M\lambda^2}{9r^3}} dr
= \frac{3r^4}{4M\lambda^2} \e^{- \frac{4M\lambda^2}{9r^3}} - \int \frac{12r^3}{4M\lambda^2} \e^{- \frac{4M\lambda^2}{9r^3}} dr \nonumber \\
=&\, \left( \frac{3r^4}{4M\lambda^2} + \mathrm{O}\left( r^7 \right) \right) \e^{- \frac{4M\lambda^2}{9r^3}} \, , \nonumber \\
\int \e^{\frac{4M\lambda^2}{9r^3}} dr
=&\, \left( - \frac{3r^4}{4M\lambda^2} + \mathrm{O}\left( r^7 \right) \right) \e^{\frac{4M\lambda^2}{9r^3}} \, .
\end{align}
we find
\begin{align}
\label{Hywrd3}
f(r) \sim - \frac{3 c_0 r^4}{4M\lambda^2} \e^{\frac{4M\lambda^2}{9r^3}}+ c_1 + \frac{9r^4}{80M\lambda^2} + \mathrm{O}\left( r^7 \right) \, .
\end{align}
If $c_0$ does not vanish, the first term in (\ref{Hywrd3}) becomes dominant and we find
\begin{align}
\label{Hywrd3B}
f''(r) \sim - \frac{9 c_0 r^4}{M\lambda^2} \e^{\frac{4M\lambda^2}{9r^3}} \, .
\end{align}
Therefore if we choose $c_0>0$, $f''(r)<0$ and the scalar field $\xi$ could not be a ghost.
The behaviors of $f$ and therefore the scalar field $\xi$ become very singular at the origin.
If we should avoid the singularity, we need to put $c_0=0$.
In the case $c_0=0$,  however, we find
\begin{align}
\label{Hywrd3C}
f''(r) \sim \frac{27r^2}{20M\lambda^2} >0 \, .
\end{align}
and therefore there appear the ghosts.

On the other hand, when $r$ is large, we obtain
\begin{align}
\label{Hywrd4}
\Upsilon(r)= - \frac{1}{3r}\,, \quad
f(r)= \int \frac{3M \lambda^2}{2}\left( \int r^{-3-\frac{4}{3}} dr+ c_0 \right)r^\frac{4}{3} dr +c_1
= \frac{9M \lambda^2}{20r} + \frac{3}{7} c_0 r^\frac{7}{3} + c_1 \, .
\end{align}
If $c_0$ does not vanish, the second term in (\ref{Hywrd4}) dominates for large $r$ and we find
\begin{align}
\label{Hywrd5}
f''(r) \sim \frac{4}{3}c_0 r^\frac{1}{3}\, .
\end{align}
By choosing $c_0<0$, $f''(r)<0$ and the scalar field $\xi$ could not be the ghost.

The summary of the above section is the fact that we have constructed the model of the scalar-Einstein-Gauss-Bonnet gravity which realizes Hayward BH in (\ref{Hayward1}).
As long as we observe the behaviors when $r\to 0$ and $r\to \infty$, the ghost might be excluded by choosing the constants of the integration
although it is highly non-trivial to check if we can choose the constants of the integration consistently.

{

We should also note that we cannot constrain the constants of the integration by any solar system or galactic test.
As mentioned after Eq.~(\ref{eq:GB}), in order to avoid the appearance of the fifth force, we have assumed that the scalar field does not directly couple with matters.
Therefore the motions of the matters are only governed by the geometry (\ref{Hayward1}),
which is that of the standard Hayward black hole, and the motions do not depend on the constants of the integration.}

\section{Ghosts for general metric}\label{S6}

Now, to investigate the problem of ghosts, we consider the small $r$ and large $r$ behaviors of the solution of Eq.~(\ref{f})
corresponding to the general form of the ansatz $a(r)$ in (\ref{met1}) and check the results for Hayward BH in the last section.
For this purpose, we consider the behavior of $f(r)$ when $r$ is small and when $r$ is large enough.

If we rewrite the ansatz $a$ as
\begin{align}
\label{deltaa}
a(r)=1 + \delta a(r)\, ,
\end{align}
and we assume that $\left| \delta a \right|\ll 1$, then Eq.~(\ref{f}) can be rewritten as:
\begin{align}
\label{Hayward7}
0 = - 8 r^3 \left( r^{-2} \delta a \right)' f'' -8 r^3 \left( r^{-2} \delta a' \right)' f'- 2 \delta a+ r^2 \delta a'' \, .
\end{align}
If we assume that $\delta a \sim a_0 r^n$ where $a_0$ is a constant and we consider the case of small $r$, i.e., $r\to 0$, then the correction will be of $\mathcal{O}\left( r^{n+1} \right)$.
On the other hand, if we consider the case $r\to \infty$, the correction is $\mathcal{O}\left( r^{n-1} \right)$.
We also note that when $n=-1, 2$, we obtain $- 2 \delta a+ r^2 \delta a''\sim 0$, therefore, we need to reconsider these two cases separately.
Moreover, if we further assume that $f \sim f_0 r^m$, we find
\begin{align}
\label{Hayward7b}
 - 8 r^3 \left( r^{-2} \delta a \right)' f'' -8 r^3 \left( r^{-2} \delta a' \right) f' \sim - 8 m \left\{ \left( n-2 \right) \left( m -1 \right) + n \left( n - 3 \right) \right\} a_0 f_0 r^{-2 + m+n}\, .
\end{align}
Therefore when $m=0$, or $m=1 - \frac{n\left( n - 3 \right)}{n-2} = - \frac{n^2 - 4n + 2}{n-2}$ if $n\neq 2$,
we find $ - 8 r^3 \left( r^{-2} \delta a \right)' f'' -8 r^3 \left( r^{-2} \delta a' \right) f' \sim 0$.
When $n\neq -1$ nor $n\neq 2$ and $m\neq 0$ nor $m\neq1 - \frac{n\left( n - 3 \right)}{n-2} = - \frac{n^2 - 4n + 2}{n-2}$, we find $m=2$.
When $n=-1$ or $n=2$, we assume
\begin{align}
\label{sub}
\delta a \sim a_0 r^n + a_1 r^l\, , \quad \left(l\neq -1, 2\right)\, .
\end{align}
Then we find the following behavior of $f$
\begin{itemize}
\item When $n\neq -1$ nor $n\neq 2$, we find $m=2$ and
\begin{align}
\label{Hayward8}
f(r) \sim f_0 + f_1 r^{- \frac{n^2 - 4n + 2}{n-2}} + \frac{\left( n + 1 \right) \left( n - 2 \right)}{16 \left\{ n^2 - 2 n -2 \right\}} r^2 \, .
\end{align}
\item When $n=-1$, we obtain $m=2+l-n=3+l$
\begin{align}
\label{Hayward9}
f(r) \sim f_0 + f_1 r^\frac{7}{3} - \frac{\left( l + 1 \right) \left( l - 2 \right)a_1}{8 \left( l + 3 \right) \left( 3 l+2 \right) a_0} r^{3+l} \, ,
\end{align}
which may tells that $l\neq -3$.
\item When $n=2$, we find $m=l$ and
\begin{align}
\label{Hayward10}
f(r) \sim f_0 - \frac{\left( l + 1 \right) \left( l - 2 \right)a_1}{16 l a_0} r^l \, .
\end{align}
We should note that when $n=2$, the second order differential equation~(\ref{Hayward7}) reduces to the first order differential equations and therefore
there appears only one constant $f_0$ of the integration.
Another constant of the integration corresponding to the original second-order differential equation may appear in a form that cannot be expanded by the power of $r$.
\end{itemize}

We now investigate the behavior when $r$ is small and find the necessary conditions to avoid the ghosts.
When $n<0$, $\delta a$ diverges in the limit $r\to 0$ and the assumption $\left| \delta a \right|\ll 1$ is violated.
This tells us that we only need to consider the case $n\geq 0$.
\begin{itemize}
\item When $n\neq 2$ and therefore $m=2$, we find that if $n>1+\sqrt{3}$ $\left( 0<n<1+\sqrt{3} \right)$,
we obtain $- \frac{n^2 - 4n + 2}{n-2}<2$ $\left( - \frac{n^2 - 4n + 2}{n-2}>2 \right)$.
We should note that because the coefficient of the third term in (\ref{Hayward8}) diverges when $n=1+\sqrt{3}$, we may assume $n\neq 1+\sqrt{3}$.
Therefore we find,
\begin{itemize}
\item When $n>1+\sqrt{3}$, the second term in (\ref{Hayward8}) dominates for $f''(r)$ and
by using (\ref{xi3}), if $f_1$ does not vanish, we find the necessary condition $\left(a - 1 \right)f''(r)>0$ to avoid ghosts is given by
\begin{align}
\label{gh1}
a_0 f_1 \left( n^2 - 4n + 2 \right) \left( n - 3 \right) n > 0 \, .
\end{align}
On the other hand, if $f_1$ vanishes, we find
\begin{align}
\label{gh1b}
\frac{\left( n + 1 \right) \left( n - 2 \right)}{16 \left\{ n^2 - 2 n -2 \right\}}>0\, ,
\end{align}
and therefore $f''(r)>0$ and therefore the condition for the absence of the ghosts is given by
\begin{align}
\label{gh1c}
a_0> 0\, .
\end{align}
\item When $0<n<1+\sqrt{3}$, the third term in (\ref{Hayward8}) dominates for $f''(r)$.
If $2<n<1+\sqrt{3}$, we find $f''(r)<0$, and therefore the condition or the absence of the ghosts is given by
\begin{align}
\label{gh1d}
a_0< 0\, .
\end{align}
On the other hnad, if $0<n<2$, we obtain $f''(r)>0$, and obtain the condition (\ref{gh1c}).
\end{itemize}
\item When $n=2$ and therefore $m=l$,
because we are considering the case that $r$ is small, we require $l>n=2$ so that the second term in (\ref{sub}) becomes less
dominant than the first term.
If we forget the term which cannot be expanded by the power of $r$, we find
\begin{align}
\label{gh2}
f''(r) \sim  - \frac{\left( l + 1 \right) l \left( l-1 \right) \left( l - 2 \right)a_1}{16 l a_0} r^{l-2} \, .
\end{align}
Because $l>2$, we find $\frac{\left( l + 1 \right) l \left( l-1 \right) \left( l - 2 \right)}{16 l}>0$ and therfore the necessary condition to avoid ghosts is
\begin{align}
\label{gh3}
a_1<0 \, .
\end{align}
\end{itemize}

In the case of Hayward BH in (\ref{Hayward1}), when $r\to 0$, we find
\begin{align}
\label{Hayward11}
a(r) \sim 1 - \frac{r^2}{\lambda^2} + \frac{r^5}{2M\lambda^4} + \mathcal{O}\left( r^8 \right) \, ,
\end{align}
which corresponds to the case that $n=2$, $l=5$, $a_0=- \frac{1}{\lambda^2}$, and $a_1 = \frac{1}{2M\lambda^4}$, therefore by using (\ref{Hayward11}),
we find
\begin{align}
\label{Hayward12}
f(r) = f_0 + \frac{9r^5}{80M\lambda^2} \, .
\end{align}
As we mentioned, another solution corresponding to the homogeneous equation, which is proportional to another constant of the integration,
could appear because the original differential equation is second-order.
The solution could not be able to be expanded by the power of $r$.
If we neglect the solution, when $r\to 0$, we find
\begin{align}
\label{Hayward13}
f''(r) \sim \frac{9 r^3}{4M\lambda^2} > 0\, ,
\end{align}
Therefore because $a<1$, we find $2(a-1) f''<0$ and Eq.~(\ref{xi3}) tells us that $\xi$ is an imaginary number and
therefore $\xi$ is a ghost.
The behavior of (\ref{Hayward12}) is consistent with the previous result in (\ref{Hywrd3}).
The behavior of (\ref{Hayward12}) coincides with the second and third terms in (\ref{Hywrd3}).
The first term including $c_0$ in (\ref{Hywrd3}) corresponds to the term which cannot be expanded by power of $r$
as we mentioned after Eq.~(\ref{Hayward12}).

On the other hand, when $r\to \infty$, we find
\begin{align}
\label{Hayward14}
a(r) \sim 1 - \frac{2M} r + \frac{4M^2\lambda^2}{r^4} + \mathcal{O}\left( r^{-7} \right) \, .
\end{align}
Therefore we find $n=-1$, $l=-4$, $a_0=-2M$, and $a_1 = 4M^2 \lambda^2$ and Eq.~(\ref{Hayward9}) gives
\begin{align}
\label{Hayward15}
f(r) \sim f_0 + f_1 r^\frac{7}{3} + \frac{9M\lambda^2}{20 r} \, .
\end{align}
When $f_1=0$, because $f''(r)>0$, $\xi$ becomes the imaginary, that is, $\xi$ should be ghost.
The behavior of (\ref{Hayward15}) is completely consistent with the previously obtained behavior in (\ref{Hywrd4}).

The summary of this section is the fact that we confirmed the results in the previous section about the asymptotic behaviors of $f(r)$ when $r$ is small or large.
Furthermore, we have considered more general cases as given in (\ref{Hayward8}), which give the necessary conditions to exclude ghosts although
the conditions are not sufficient.
Because the behavior of Hayward black hole in (\ref{Hayward13}) seems to imply the existence of the ghosts at least when $r$ is small,
we may modify the original metric in (\ref{Hayward1}) so that the obtained necessary condition is satisfied.

\section{Geodesic deviation}\label{S66}

In this section, we study the stability condition of a particle motion in the background of Hayward BH using the geodesic deviation.
This is an important test to know where the region of the Hayward BH becomes stable.
It is well known that the geodesics of a particle in a background of a gravitational field are described by
\begin{align}
\label{ge}
0=\frac{d^2 x^\sigma}{d\varepsilon^2}+ \left\{\begin{array}{c} \sigma \\ \mu \nu \end{array} \right\}
\frac{d x^\mu}{d\varepsilon} \frac{d x^\nu}{d\varepsilon}\, ,
\end{align}
with $\varepsilon$ being the affine parameter along the trajectory.
Eq.~(\ref{ge}) is the trajectory equation where their deviation yields the following form \cite{dInverno:1992gxs},
\begin{align}
\label{ged}
0= \frac{d^2 \eta^\sigma}{d\varepsilon^2}+ 2\left\{\begin{array}{c} \sigma \\ \mu \nu \end{array} \right\}
\frac{d x^\mu}{d\varepsilon} \frac{d \eta^\nu}{d\varepsilon} + \left\{\begin{array}{c} \sigma \\ \mu \nu \end{array} \right\}_{,\rho}
\frac{d x^\mu}{d\varepsilon} \frac{d x^\nu}{d\varepsilon}\eta^\rho\, ,
\end{align}
with $\eta^\rho$ being the deviation 4-vector.
The use of Eqs.~(\ref{ge}) and (\ref{ged}) into the line-element (\ref{Hayward1}) yields
\begin{align}
\label{gedi}
0=\frac{d^2 t}{d\varepsilon^2}\, , \quad
0=\frac{1}{2} a'(r)\left(\frac{d t}{d\varepsilon}\right)^2-r\left(\frac{d \phi}{d\varepsilon}\right)^2\,, \quad
0=\frac{d^2 \theta}{d\varepsilon^2}\, ,\quad
0=\frac{d^2 \phi}{d\varepsilon^2}\, ,
\end{align}
and
\begin{align}
\label{ged1}
0=&\, \frac{d^2 \eta^1}{d\varepsilon^2}+a(r)a'(r) \frac{dt}{d\varepsilon} \frac{d \eta^0}{d\varepsilon}-2r a(r) \frac{d \phi}{d\varepsilon} \frac{d \eta^3}{d\varepsilon} \nonumber \\
&\, +\left[\frac{1}{2}\left(a'^2(r)+a(r) a''(r) \right)\left(\frac{dt}{d\varepsilon}\right)^2-\left(a(r)+ra'(r) \right) \left(\frac{d\phi}{d\varepsilon}\right)^2 \right]\eta^1\, , \nonumber\\
0=&\, \frac{d^2 \eta^0}{d\varepsilon^2} + \frac{a'(r)}{a(r)}\frac{dt}{d\varepsilon} \frac{d \zeta^1}{d\varepsilon}\, ,\quad
0= \frac{d^2 \eta^2}{d\varepsilon^2}+\left( \frac{d\phi}{d\varepsilon}\right)^2 \eta^2\, , \quad
0= \frac{d^2 \eta^3}{d\varepsilon^2} + \frac{2}{r} \frac{d\phi}{d\varepsilon} \frac{d \eta^1}{d\varepsilon}\, ,
\end{align}
where $a(r)$ is the ansatz defined by the Eq.~(\ref{Hayward1}). Eqs.~(\ref{gedi}) and (\ref{ged1}) are the equations for
the geodesics and geodesic deviations, respectively.
The use of the circular orbit
\begin{align}
\theta=\frac{\pi}{2}\, , \quad
0= \frac{d\theta}{d\varepsilon}\, , \quad 0= \frac{d r}{d\varepsilon}\, ,
\end{align}
yield
\begin{align}
\left(\frac{d\phi}{d\varepsilon}\right)^2= \frac{a'(r)}{r \left(2a(r)-ra'(r) \right)}\, , \quad \left( \frac{dt}{d\varepsilon}\right)^2= \frac{2}{2a(r)-ra'(r)}\, .
\end{align}

\begin{figure}
\centering
\includegraphics[scale=0.3]{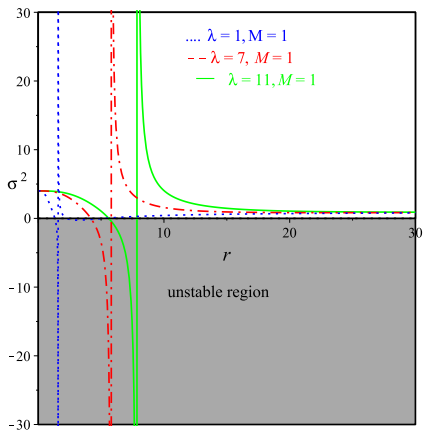}
\caption{Schematic plot of the radial coordinate $r$ vs. $\sigma^2$ which shows that stability condition given by Eq.~(\ref{stc}).}
\label{Fig:2}
\end{figure}

Eq.~(\ref{ged1}) can be reexpressed as
\begin{align}
\label{ged2}
0=&\, \frac{d^2 \eta^1}{d\phi^2}+a(r)a'(r) \frac{dt}{d\phi} \frac{d \eta^0}{d\phi}-2r a(r) \frac{d \eta^3}{d\phi}
+\left[\frac{1}{2}\left(a'(r)a'(r)+a(r) a''(r)
\right)\left( \frac{dt}{d\phi}\right)^2-\left(a(r)+ra'(r)\right) \right]\eta^1\, , \nonumber\\
0=&\, \frac{d^2 \eta^2}{d\phi^2}+\eta^2\, , \quad
0= \frac{d^2 \eta^0}{d\phi^2} + \frac{a'(r)}{a(r)} \frac{dt}{d\phi} \frac{d \eta^1}{d\phi}\, , \quad
0= \frac{d^2 \eta^3}{d\phi^2} + \frac{2}{r}\frac {d \eta^1}{d\phi}\, .
\end{align}
The second equation of Eq.~(\ref{ged2}) indicates that we possess a simple harmonic motion which means that we have a stable motion.
We can suppose the solution of the remaining of Eq. (\ref{ged2}) in the form
\begin{align}
\label{ged3}
\eta^0 = \zeta_1 \e^{i \sigma \phi}\, , \quad \eta^1= \zeta_2\e^{i \sigma \phi}\, , \quad \mbox{and} \quad
\eta^3 = \zeta_3 \e^{i \sigma \phi}\, ,
\end{align}
where
$\zeta_1$, $\zeta_2$ and $\zeta_3$ are constants and $\phi$ should be fixed.
Substituting (\ref{ged3}) into Eq.~(\ref{ged2}), we obtain
\begin{align}
\label{con1}
\frac{3aa'-\omega^2a'-ara'^2+raa''}{a'}>0\, ,
\end{align}
which is the stability condition.
Eq.~(\ref{con1}) for the BH (\ref{Hayward1}) can be reexpressed as
\begin{align}
\label{stc}
r^6+ 22r^3M\lambda^2-32M^2\lambda^4-6Mr^5>0\, ,
\end{align}
which is the condition of stability for the BH solution~(\ref{Hayward1}) and when $\lambda=0$, we get $r>6M$,
which is the condition of stability of Schwarzschild BH solution~\cite{Misner:1973prb}.
We draw the stability condition in Figure~\ref{Fig:2}, where we have shown the region of stability for different values of $\lambda$.

\section{Summary and discussions}\label{S7}

In this study and in the framework of the scalar-Einstein-Gauss-Bonnet gravity whose action is given by Eq.~(\ref{g2}), we have constructed a model which realizes Hayward BH.
Moreover, we have shown the possibility to avoid ghosts as given in Eqs.~(\ref{Hywrd3B}) and (\ref{Hywrd5}).
These equations show that if we can choose the constants of integration so that the arbitrary function appears in Eq.~(\ref{g2}), i.e., $f''(r)<0$, then the ghosts may disappear.
We have only shown that we can obtain $f''(r)<0$ at least when $r$ is small or when $r$ is large.
It is not clear if we can choose the constants of the integration so that $f''(r)<0$ simultaneously when $r$ is small or when $r$ is large.
We also need to show that $f''(r)<0$ in all the regions of $r$ to show the absence of ghosts.
Anyway, our model could not be easily excluded due to the ghosts.
We also need to show the stability of the solution under the perturbation.

Therefore there remain many difficulties in the model however, we have shown that the problem of information loss might be solved by
introducing stringy corrections like the Gauss-Bonnet invariant.

\end{document}